\documentclass[a4paper,11pt]{amsart}
\begin{document}
\hyphenation{gra-vi-ta-tio-nal re-la-ti-vi-ty Gaus-sian
re-fe-ren-ce re-la-ti-ve gra-vi-ta-tion Schwarz-schild
ac-cor-dingly gra-vi-ta-tio-nal-ly re-la-ti-vi-stic pro-du-cing
de-ri-va-ti-ve ge-ne-ral ex-pli-citly des-cri-bed ma-the-ma-ti-cal
de-si-gnan-do-si coe-ren-za pro-blem gra-vi-ta-ting geo-de-sic
per-ga-mon cos-mo-lo-gi-cal gra-vity cor-res-pon-ding
de-fi-ni-tion phy-si-ka-li-schen ma-the-ma-ti-sches ge-ra-de
Sze-keres con-si-de-red tra-vel-ling ma-ni-fold re-fe-ren-ces}
\title[On the relation between Schwarzschild's and Kerr's manifolds]
{{\bf On the relation between Schwarzschild's\\and Kerr's
manifolds}}

\author[Angelo Loinger]{Angelo Loinger}
\address{A.L. -- Dipartimento di Fisica, Universit\`a di Milano, Via
Celoria, 16 - 20133 Milano (Italy)}
\author[Tiziana Marsico]{Tiziana Marsico}
%\date{}
\address{T.M. -- Liceo Classico ``G. Berchet'', Via della Commenda, 26 - 20122 Milano (Italy)}
\email{angelo.loinger@mi.infn.it} \email{martiz64@libero.it}
%\thanks{}

\vskip0.50cm

\begin{abstract}
Kerr's manifold is only a Schwarzschild's manifold as ``seen'' by
a suitably rotating coordinate system. By taking into account this
fact, Kerr's manifold can be ``reduced'' to a Schwarzschild's
manifold. -- In a final \emph{aper\c cu} we summarize the main
steps of our reasoning.
\end{abstract}

\maketitle

%%\begin{equation} \label{eq:sevenprime}
%%    \ddot{\Re} + \f\textbf{5}. External
%% {\kappa}{6}\Re \rho=0 , \tag{7'}
%% \end{equation}
%% ``mechanisms'' \textrm{d} \`a
%% \cite{1}
%% eqs.(\ref{eq:six})
%% Schwarzschild


\normalsize

\vskip1.20cm \noindent \textbf{1.} -- The standard
(Hilbert-Droste-Weyl) form for the $\textrm{d}s^{2}$ of
Schwarzschild's manifold of a gravitating point mass $m$ is -- if
$r'$, $\vartheta'$, $\varphi'$ are spherical polar coordinates:

\begin{eqnarray} \label{eq:one}
\textrm{d}s^{2} & = & \left(1-\frac{2m}{r'}\right)^{-1}
\textrm{d}r'^{2} + r'^{2}\left(\textrm{d}\vartheta'^{2} +
\sin^{2}\vartheta' \textrm{d}\varphi'^{2}\right) -
\nonumber\\
& & {} %%
- \left(1-\frac{2m}{r'} \right) \textrm{d}t'^{2} \quad; \quad
(G=c=1) \quad.
\end{eqnarray}

With Boyer and Lindquist (see \cite{1}, \cite{2}), the
$\textrm{d}s^{2}$ of Kerr's manifold can be written as follows:

\begin{eqnarray} \label{eq:two}
\textrm{d}s^{2} & = &
\left(\frac{\textrm{d}r^{2}}{\Delta}+\textrm{d}\vartheta^{2}\right)
\Sigma + (r^{2}+a^{2}) \sin^{2}\vartheta \, \textrm{d}\varphi^{2}
- \textrm{d}t^{2} +
\nonumber\\
& & {} %%
+ \frac{2mr}{\Sigma} \, \left(a \sin^{2}\vartheta \,
\textrm{d}\varphi + \textrm{d}t\right)^{2}  \quad; \quad (G=c=1)
\quad,
\end{eqnarray}

where: $\Sigma \equiv r^{2}+a^{2}\cos^{2}\vartheta$; $\Delta
\equiv r^{2}-2mr+a^{2}$. The parameter $\emph{a}$ has a
geometrical and kinematical meaning. \emph{The case} $m\geq a$
\emph{is physically interesting}. When $a=0$, eq. (\ref{eq:two})
coincides with eq. (\ref{eq:one}).

\vskip1.20cm \noindent \textbf{2.} -- The potential $g_{jk}$,
$(j,k=1, 2, 3, 4)$, of eq. (\ref{eq:two}) is referred to a frame
which rotates with the following angular velocity
$g_{t\varphi}/g_{\varphi\varphi}\equiv \omega$:

\begin{equation} \label{eq:twoprime}
\omega = \frac{2\, m\, a\,
r}{(r^{2}+a^{2})(r^{2}+a^{2}\cos^{2}\vartheta) + 2\, m\, a^{2}r
\sin^{2}\vartheta} \quad; \tag{2'}
\end{equation}

at $r=0$, we have $\omega=0$: this means, strictly speaking, that
the gravitating point mass does \emph{not} rotate. For $r\neq 0$
and $m \neq 0$, $\omega$ is equal to zero if and only if the
parameter $a$ is equal to zero, and in this case Kerr's manifold
coincides with Schwarzschild's manifold, as we have seen in
sect.\textbf{1}. -- Of course, an $\omega\neq 0$ generates
\emph{dragging} forces.

\vskip1.20cm \noindent \textbf{3.} -- Kerr's surface
$r=m+(m^{2}-a^{2}\cos^{2}\vartheta)^{1/2}$, which is tangent to
the surface $r=m+(m^{2}-a^{2})^{1/2}$ at $\vartheta=0$ and
$\vartheta=\pi$, has this property: if we make in eq.
(\ref{eq:two}) the coordinate shift (a legitimate choice of a new
radial coordinate):

\setcounter{equation}{2}
\begin{equation} \label{eq:three}
r \rightarrow  r+ m + (m^{2}-a^{2}\cos^{2}\vartheta)^{1/2}\quad,
\end{equation}

we obtain a $\Delta\geq 0$; the equality holds for $r=0$ and
$\cos^{2}\vartheta=1$. Indeed, the new $\Delta$ is:

\begin{eqnarray} \label{eq:four}
\Delta & = & \left[\,r+ (m^{2}-a^{2}\cos^{2}\vartheta)^{1/2} -
(m^{2}-a^{2})^{1/2} \right] \cdot \nonumber\\
& & {} %%
\cdot  \left[\,r+ (m^{2}-a^{2}\cos^{2}\vartheta)^{1/2} +
(m^{2}-a^{2})^{1/2} \right] \geq 0 \quad;
\end{eqnarray}

and if $\cos^{2}\vartheta=1$:

\begin{equation} \label{eq:fourprime}  \tag{4'}
\Delta_{\vartheta=0,\pi} = \, r \cdot \left[\,r+ 2\,
(m^{2}-a^{2})^{1/2} \right] \quad.
\end{equation}

When $a=0$, transformation (\ref{eq:three}) becomes:

\begin{equation} \label{eq:threeprime}
r \rightarrow r + 2m \quad, \tag{3'}
\end{equation}

which coincides with Brillouin (-Schwarzschild) transformation of
radial coordinate $r'$ of eq. (\ref{eq:one}) \cite{3}.

\par Brillouin's form of Schwarzschildian $\textrm{d}s^{2}$  and
the above new form of Kerr's $\textrm{d}s^{2}$  have \emph{both} a
\textbf{\emph{sole}} (and ``soft'') singularity at $r'=r=0$. This
means that eq. (\ref{eq:one}) and eq. (\ref{eq:two}) have a
physical (and mathematical) meaning only when $r'>2m$ and
$r>m+(m^{2}-a^{2}\cos^{2}\vartheta)^{1/2}$, respectively. Our
paper quoted in \cite{2} gives a striking proof of this assertion.
(Accordingly, there is no room for the physical existence of
BH's). -- Remark that the new form (\emph{\`a la} Brillouin) of
Kerr's metric is \textbf{\emph{diffeomorphic}} to the
\emph{exterior} part (\emph{i.e.}, for
$r>m+(m^{2}-a^{2}\cos^{2}\vartheta)^{1/2}$) of the form of eq.
(\ref{eq:two}), and is \emph{maximally extended}. All the
\emph{observational} results concern only the exterior part of eq.
(\ref{eq:two}), or equivalently the $r>0$ region of the new form
of $\textrm{d}s^{2}$.

 \vskip1.20cm \noindent \textbf{4.} -- We have seen that the
 singular surfaces $r'=2m$ and $r=m+(m^{2}-a^{2}\cos^{2}\vartheta)^{1/2}$ are in
 a strict correspondence. Moreover, we prove that this \emph{Kerr's singular surface can be transformed into the
 surface} $r'=2m$ \emph{with an appropriate change of general
 coordinates}.

\par From the standpoint of a three-dimensional Euclidean
\emph{Bildraum}, a ``vertical'' section of the surface
$r=m+(m^{2}-a^{2})^{1/2}$ is an ellipse (see Appendix);
accordingly, this surface is a rotational ellipsoid,
\emph{\emph{i.e.}} an oblate spheroid. And the surface $r'=2m$ is
a sphere.

\par The semi-axes, say $\alpha$ and $\beta$, of the above ellipse
are:

\begin{displaymath} \label{eq:five}
\left\{ \begin{array}{l}
\alpha=2m \quad; \quad \beta=m+(m^{2}-a^{2})^{1/2} \quad; \quad \Rightarrow \\
\alpha^{2} - \beta^{2} = \, a^{2}+2m \, \left[\,
m-(m^{2}-a^{2})^{1/2}\right] \quad. \tag{5}
\end{array} \right.
\end{displaymath}

We remark that

\setcounter{equation}{5}
\begin{equation} \label{eq:six}
\{a=0\}  \, \Leftrightarrow \, \{\gamma^{2}\equiv \alpha^{2}-
\beta^{2}=0\} \quad.
\end{equation}

Obviously, if $\xi$, $\eta$ are Cartesian orthogonal coordinates,
the equation of our ellipse can be also written as follows:

\begin{equation} \label{eq:seven}
\frac{\xi^{2}}{\alpha^{2}} \, + \, \frac{\eta^{2}}{\beta^{2}} \, =
\,  1 \quad.
\end{equation}

If $\xi=\xi'$, $\eta=(\beta / \alpha) \, \eta'$, we have

\begin{equation} \label{eq:eight}
\xi'\,^{2} + \eta'\,^{2} = (2m)^{2} \quad,
\end{equation}

where:

\begin{equation} \label{eq:nine}
\frac{\beta^{2}}{\alpha^{2}} = \frac{2\,m \, [\,
m+(m^{2}-a^{2})^{1/2}]-a^{2}}{4\,m^{2}} ^{}\quad.
\end{equation}

The transformation of the spheroid
$r=m+(m^{2}-a^{2}\cos^{2}\vartheta)^{1/2}$ into the sphere $r'=2m$
is an immediate consequence of eq. (\ref{eq:eight}).

\par This result has been obtained with a simple application of
the following theorem of projective geometry: if we have an
ellipsoid (resp. an ellipse) and a sphere (resp. a circle), there
exist \emph{collineations} that transform the ellipsoid (resp. the
ellipse) into the sphere (resp. the circle).

\par The parameter $a$ (its geometrical meaning is explained by
eq. (\ref{eq:six})), which is responsible for the spinning of
Kerr's frame, has been ``incorporated'' in new coordinates; this
implies that Kerr's manifold is \emph{not} substantially distinct
from Schwarzschild's manifold, because \emph{the ``soft''
singularities} $r'=2m$ \emph{and}
$r=m+(m^{2}-a^{2}\cos^{2}\vartheta)^{1/2}$ -- \emph{which
represent projectively the} \textbf{\emph{same}} \emph{geometrical
object} -- \emph{characterize completely these manifolds}. Indeed,
in the shifted coordinates \emph{\`a la} Brillouin
(-Schwarzschild) both manifolds are solutions of Einstein
equations $R_{jk}=0$, $(j,k=1,2,3,4)$, with \emph{one and only}
``soft'' singularity at the origin of the radial coordinates
$(r'=r=0)$; the surfaces $r=2m$ and
$r=m+(m^{2}-a^{2}\cos^{2}\vartheta)^{1/2}$ are really \emph{not}
surfaces but \emph{single points}.

\vskip1.20cm \noindent \textbf{5.} -- The above conclusion is
similar to this Weyl's result \cite{4}: the $\textrm{d}s^{2}$ of
eq. (\ref{eq:one}) can be expressed also in a cylindrical system
of coordinates (Weyl's ``canonical'' system) $z^{*}, r^{*},
\vartheta^{*}$. Then, the ``globe'' $r'=2m$ becomes the
``segment'' $-m\leq z^{*}\leq +m$.

\par Our case is a little more involved, owing to the presence of
the parameter $a$, which however can be ``taken up'' by the
coordinate change that allows the transformation of the spheroid
$r=m+(m^{2}-a^{2}\cos^{2}\vartheta)^{1/2}$ into the sphere
$r'=2m$.

\par When $m=0$ and $a\neq 0$, there is a simple relation between
$(r', \vartheta', \varphi', t')$ and $(r, \vartheta, \varphi, t)$,
see Appendix of paper \cite{2}; in this case, eq. (\ref{eq:one})
and eq. (\ref{eq:two}) give only two different forms of
\emph{Minkowski} interval $\textrm{d}s_{M}^{2}$.

\vskip1.20cm \noindent \textbf{6.} -- A consideration on the role
of the \emph{Killing vectors} \cite{5}. As it is well known, they
yield an \emph{invariant} description of the symmetry properties
of a given manifold. However, it is necessary to distinguish the
Riemann-Einstein manifolds generated by \emph{extended} material
distributions from the Riemann-Einstein manifolds generated by
\emph{punctual} masses, that are solutions of $R_{jk}=0$ with a
singularity at the origin of the coordinates. In this second case,
we can have manifolds with \emph{any} kind of symmetry: indeed, a
mass point can be considered as a kind  of limit of a material
distribution of any symmetry -- and therefore a coordinate system
``adapted''  to a chosen symmetry is also adequate to the field
generated by the material point. In other terms, the manifold
created by a mass point does \emph{not} possess a definite
symmetry \emph{of its own}. In sect.\textbf{5} we have mentioned
an example given by Weyl \cite{4}. Another example is Kerr's
manifold: as we have seen, Kerr's oblate spheroid
$r=m+(m^{2}-a^{2}\cos^{2}\vartheta)^{1/2}$ can be transformed,
with a simple collineation, into the sphere $r'=2m$. In this way,
\emph{Kerr's manifold is ``reduced'' to Schwarzschild's manifold
of a mass point at rest}. (\emph{N.B.} -- The use of a
\emph{Bildraum} for the proof of this ``reduction'' does not
restrict the validity of our result).

\par The $\textrm{d}s^{2}$ of eq. (\ref{eq:one}) can be considered,
\emph{e.g.}, as the limit of the $\textrm{d}s^{2}$ of a
homogeneous sphere of an incompressible fluid, whose radius goes
to zero \cite{6}. The $\textrm{d}s^{2}$ of eq. (\ref{eq:two}) can
be considered, \emph{e.g.}, as the limit of the $\textrm{d}s^{2}$
of the above contracting sphere as ``viewed'' by a frame which
rotates with the angular velocity $\omega$ of eq.
(\ref{eq:twoprime}).

\par Summing up, the difference between Kerr's manifold and
Schwarzschild's manifold is \emph{only} a difference of reference
systems: Kerr's  metric is described by a \emph{rotating} frame,
Schwarzschild's metric by a \emph{static} frame. In \emph{both}
cases the material agent is the \emph{same}: a point mass.

\vskip1.20cm \noindent \textbf{7.} -- \emph{\textbf{Aper\c cu}}.
-- \emph{i}) If the angular velocity of rotation contained in
Kerr's metric is equal to zero, Kerr's manifold coincides with
Schwarzschild's manifold created by a point mass at rest. --
\emph{ii}) The coordinate shift $r \rightarrow
r+m+(m^{2}-a^{2}\cos^{2}\vartheta)^{1/2}$ in Kerr's metric gives
an expression of the $\textrm{d}s^{2}$  with a \emph{sole} (and
``soft'') singularity at $r=0$. -- \emph{iii}) The oblate spheroid
$r=m+(m^{2}-a^{2}\cos^{2}\vartheta)^{1/2}$ can be transformed into
the sphere $r'=2m$. -- \emph{iv}) The difference between Kerr's
metric and Schwarzschild's metric rests \emph{only} on the
difference between the respective reference frames. Accordingly,
Kerr's metric can be ``reduced'' to Schwarzschild's metric  by
virtue of result \emph{iii}). -- \emph{v}) Of course, Kerr's
potential $g_{jk}$ gives origin to a Thirring-Lense
\emph{dragging} effect. However, all dragging forces are
\emph{only} caused by the chosen reference frames, and therefore
do not have an \emph{invariant} character. -- \emph{vi}) All
\emph{observational} data are in accord with our analysis. --

% \newpage
\vskip2.00cm
\begin{center}
\noindent \small \emph{\textbf{APPENDIX}}
\end{center}
\normalsize \noindent \vskip0.80cm

\par We give here the banal proof that the vertical section of
Kerr's surface $r=m+(m^{2}-a^{2}\cos^{2}\vartheta)^{1/2}$, where
$0\leq r< +\infty$ and $0\leq \vartheta \leq \pi$ , is an ellipse.

\par If $\alpha$, $\beta$ are the semi-axes of a generic ellipse,
and $\chi$, ($0\leq \chi < 2\pi$), is Kepler's eccentric anomaly,
our curve can be described by the following equations -- as it is
well known:

\begin{equation} \label{eq:A1}
\xi = \alpha \cos\chi \quad; \quad \eta = \beta \sin \chi \quad,
\tag{A1}
\end{equation}

where $\xi$, $\eta$ are Cartesian orthogonal coordinates. If
$\varrho^{2} =  \xi^{2} + \eta^{2}$, we have:

\begin{equation} \label{eq:A2}
\varrho = \left[ \beta^{2} - (\beta^{2}-\alpha^{2}) \cos^{2} \chi
\right]^{1/2} \quad. \tag{A2}
\end{equation}

Now, the equation $r-m=(m^{2}-a^{2}\cos^{2}\vartheta)^{1/2}$ can
also represent the half ($\vartheta$ is a colatitude) of the
vertical section of the above Kerr's surface. We see that:

\begin{equation} \label{eq:A3}
\varrho\,(\chi=0) = \alpha \quad; \quad \varrho\,(\chi=\pi/2) =
\beta \quad, \tag{A3}
\end{equation}

\begin{equation} \label{eq:A4}
r\,(\vartheta=\pi/2) = 2m \quad; \quad r\,(\vartheta=0) =
m+(m^{2}-a^{2})^{1/2} \quad, \tag{A4}
\end{equation}

\qquad \qquad \qquad \qquad \qquad \qquad \qquad \qquad \qquad
\qquad \qquad \qquad \qquad \qquad \emph{Q.e.d.}

\vskip0.80cm Accordingly, by virtue of the axial symmetry of
Kerr's $\textrm{d}s^{2}$, the surface
$r=m+(m^{2}-a^{2}\cos^{2}\vartheta)^{1/2}$ is a rotational
ellipsoid. --

\end{document}